\documentclass[usenatbib]{pasa}%
\usepackage{upgreek}
\usepackage{amssymb,amsmath}
\usepackage{graphicx}
\usepackage{multirow}
\usepackage[authoryear]{natbib}
\usepackage{tabularx}
\usepackage{subfigure}
\usepackage[flushleft]{threeparttable}
\usepackage[T1]{fontenc}

\def\cite{\citep}

\title[FRB catalogue]{FRBCAT: The Fast Radio Burst Catalogue}
\author[Petroff et al.]{E. Petroff$^{1,2,3}$\thanks{ebpetroff@gmail.com}, E. D. Barr$^{1,3}$, A. Jameson$^1$, E. F. Keane$^{4,3,1}$, M. Bailes$^{1,3}$, M. Kramer$^{5,6}$, V. Morello$^{1,3}$, D. Tabbara$^{1}$ \and W. van Straten$^{1,3}$ \\
\affil{$^1$Swinburne University of Technology, Swinburne University of Technology, P.O. Box 218, Hawthorn, VIC 3122, Australia}%
\affil{$^2$CSIRO Astronomy \& Space Science, Australia Telescope National Facility, P.O. Box 76, Epping, NSW 1710, Australia}
\affil{$^3$ARC Centre of Excellence for All-sky Astrophysics (CAASTRO)}
\affil{$^4$SKA Organisation, Jodrell Bank Observatory, Cheshire, SK11 9DL, UK}
\affil{$^5$Max Planck Institut f\"{u}r Radioastronomie, Auf dem H\"{u}gel 69, D-53121 Bonn, Germany}
\affil{$^6$Jodrell Bank Centre for Astrophysics, University of Manchester, Alan Turing Building, Oxford Road, Manchester M13 9PL, United Kingdom}
}%
\jid{PASA}
\doi{10.1017/pas.\the\year.xxx}
\jyear{\the\year}

\bibpunct{(}{)}{;}{a}{}{,}
\begin{document}%
\begin{abstract}

Here we present a catalogue of known Fast Radio Burst (FRB) sources in the form of an online catalogue, FRBCAT. The catalogue includes information about the instrumentation used for the observations for each detected burst, the measured quantities from each observation, and model-dependent quantities derived from observed quantities. To aid in consistent comparisons of burst properties such as width and signal-to-noise ratios we have reprocessed all the bursts for which we have access to the raw data, with software which we make available. The originally derived properties are also listed for comparison. The catalogue is hosted online as a \textsc{MySQL} database which can also be downloaded in tabular or plain text format for off-line use. This database will be maintained for use by the community for studies of the FRB population as it grows. 

\end{abstract}
\maketitle%
\section{INTRODUCTION }
\label{sec:intro}

Fast radio bursts (FRBs) -- bright millisecond radio pulses -- have generated considerable excitement within the astronomy community, with potential uses in cosmology, the intergalactic and interstellar media, coherent emission processes, compact objects, and more \citep{SKATransients}. To date, the majority of FRB discoveries have been made at the Parkes radio telescope \citep{Lorimer07,Keane12,Thornton13,SarahFRB,PetroffFRB,Ravi2015,Champion2015}, with additional discoveries made at the Arecibo and Green Bank telescopes \citep{Spitler14,GBTBurst}. With mounting evidence for their astrophysical origin ever more projects dedicated to FRB searches are arising at new and existing telescopes \citep{Law15,Karastergiou2015,Tingay2015,alfaburst}. At present, the rate of discovery is still rather slow, but new facilities are under construction which promise an explosion in FRB discovery rates \citep{SKATransients}. It is therefore timely to catalogue what we know so far and to re-measure quantities for the published FRBs in a uniform and systematic fashion. Furthermore, it is important to identify commonly used derived parameters which are model-dependent and examine the various degrees to which these are uncertain.

Our catalogue is partially presented here for the currently published FRBs in tabular form but is also fully available in the form of an online catalogue and database at \texttt{http://www.astronomy.swin.edu.au/pulsar/frbcat/} for community use. Where data were publicly available, we have performed a systematic re-analysis with software using the methods outlined in this work. 

In the following sections we describe the online interface and how to use the catalogue. In Section~\ref{sec:quantities} we present the entries in the catalogue used to describe each burst. Section~\ref{sec:howto} provides information on how to use the catalogue; we conclude in Section~\ref{sec:conclusions}.

\section{PARAMETERS}
\label{sec:quantities}

A list is presented in the following subsections of all the quantities and parameters displayed in the catalogue. These have been divided into three categories: observation parameters, observed parameters, and derived parameters. Observation parameters are related to the telescope and instrument used for observation. Observed parameters relate to the detected burst and quantities obtained through direct processing of the data. Derived parameters related to cosmology and distance are based upon combining the observed parameters with model-dependent numbers for cosmological values ($\Omega_\mathrm{M}$, $\Omega_\Lambda$, $H_0$) and the ionised component of the Galaxy (DM$_\mathrm{Galaxy}$). 

The separation of these three categories is for several reasons. Firstly, they have been separated to draw clear distinctions between quantities that are directly measured, speculative based on FRB position in the telescope beam, and speculative due to a combination of the positional uncertainty and the uncertainty of a model or a models. Errors are provided for measured quantities. Secondly, this catalogue is intended for use as the FRB population continues to grow.
Each observation will also have its own set of observed and derived parameters which will be attached to it. Additional sets of derived parameters can be added when data are re-analysed in new ways.

Observed parameters are included with a reference to the measurements leading to the numbers presented. In the cases where the data are publicly available two (or more) sets of observed parameters for the original detection observation are included: those published in the discovery paper and those obtained through our systematic reanalysis described here. A reference is included to make note of the measurement method. 

The signal to noise (S/N), width, and dispersion measure (DM) for each burst are re-derived using the \textsc{destroy}\footnote{https://github.com/evanocathain/destroy\_gutted} single pulse search code, the \textsc{psrchive}\footnote{http://psrchive.sourceforge.net/} package and scripts made publicly available through the FRBCAT github page\footnote{https://github.com/frbcat/FRBCAT\_analysis}. These numbers have in turn been independently verified using a python-based fitting routine.
For the python-based approach a filterbank data block of a few seconds around the event was extracted. To remove the effects of narrowband interference, frequency channels with abnormally high or low variance over the data segment were masked. We used the outlier detection method known as Tukey's rule for outliers \citep[see e.g.][]{Chandola}. The resultant data block was then de-dispersed at different trial DM values using the smallest DM trial step allowed by the original time resolution of the raw data. In the case of Parkes search-mode data since 2008, $\Delta \mathrm{DM}$ = 0.0488.

Every de-dispersed time series was then separately searched for pulses. A running median of length 0.5 seconds was subtracted to mitigate the effects of low-frequency noise, then the time series was normalized to zero mean and unit variance by carefully excluding outlier values (i.e. time samples containing the FRB signal). Finally, a standard pulse search algorithm was used, which consists in convolving the time series with a set of top-hat pulse templates \citep{PulsarHandbook}, here with widths covering all values from from 1 to 400 time samples, trying to maximize the response:

\begin{equation}
S/N = \frac{1}{\sqrt{W}} \sum_{j=i}^{i+W} t_j
\end{equation}

where $W$ is the width in number of bins of the top-hat pulse template, $i$ the trial starting sample index of the pulse, and $t_j$ the j-th bin of the time series. The maximum S/N value points to the optimal FRB parameters: dispersion measure, pulse width, and arrival time.

Both the published and re-derived parameters are included on the page for an individual burst. If available, the re-derived values are the ones presented on the catalogue homepage to maintain consistency across the sample where possible, as some search codes used for the initial discoveries have been found to under-report signal to noise \citep{KeanePetroff}.

Values for some of the following parameters for the published FRBs are presented in Table~\ref{tab:tab1}. Values for cosmological parameters, as derived by \textsc{cosmocalc} \citep{CosmologyCalc}, are also included in the catalogue.




\subsection{Observation parameters}

\noindent \textsc{\textbf{Telescope}} --- Telescope used to take the observation.

\noindent \textsc{\textbf{Receiver}} --- Receiver system on the telescope used to take observations. At Parkes the primary instrument is the 13-beam multibeam receiver (MB20; \citeauthor{multibeam}, \citeyear{multibeam}), at Arecibo the 7-beam ALFA receiver \citep{Spitler14}, and at the Green Bank Telescope (GBT) the 800 MHz receiver~\cite{GBTBurst}.

\noindent \textsc{\textbf{Backend}} --- The data recording system used for the observation. At Parkes two primary data recording systems have been used: the analogue filterbank (AFB) and Berkeley Parkes Swinburne Recorder (BPSR; \citeauthor{Keith10}, \citeyear{Keith10}). The Mock spectrometers and the GUPPI backends have been used at Arecibo and the GBT, respectively.

\noindent \textsc{\textbf{Beam}} --- The telescope beam number in which the FRB was detected. For FRBs detected with single-pixel feeds this value is set to 1.

\noindent \textsc{\textbf{RAJ}} --- The right ascension in J2000 coordinates of the pointing centre of the detection beam. This is not the location of the burst and corresponds only to the position of the centre of the telescope beam. Error on pointing accuracy of the telescope is also listed.

\noindent \textsc{\textbf{DECJ}} --- The declination in J2000 coordinates of the pointing centre of the detection beam. As with RAJ, this is not the location of the burst but only the position of the centre of the beam. Error on pointing accuracy of the telescope is also listed. 

\noindent \textit{\textbf{gl}} --- The Galactic longitude, in degrees, of the pointing centre of the beam. 

\noindent \textit{\textbf{gb}} --- The Galactic latitude, in degrees, of the pointing centre of the beam.

\noindent \textsc{\textbf{Beam FWHM}} --- The diameter of the full width half maximum of the primary lobe of the detection beam in arcminutes. This parameter serves as the best estimate of the positional uncertainty of the burst as a detection is most likely to lie within this area. 

\noindent \textsc{\textbf{Sampling time}} --- The length of a time sample for the observation, in milliseconds. Sampling times between 1 ms and 64 $\upmu$s are common for observations in which FRBs were detected.

\noindent \textsc{\textbf{Bandwidth}} --- The observing bandwidth in MHz. In the case of observations using BPSR, the system bandwidth is 400 MHz but the usable bandwidth (after excision of channels rendered unusable by interference) is 338.281 MHz; the latter is quoted in this catalogue as it is the relevant number for calculations of signal to noise and flux density.

\noindent \textsc{\textbf{Centre frequency}} --- The centre frequency of the observation in MHz. 

\noindent \textsc{\textbf{Number of polarisations}} --- The number of polarisations used to record the total signal. 

\noindent \textsc{\textbf{Channel bandwidth}} --- The bandwidth of the individual frequency channels in MHz.

\noindent \textsc{\textbf{Bits per sample}} --- The number of bits recorded for an individual time sample in the final data product. 

\noindent \textsc{\textbf{Gain}} --- The telescope gain in units of K Jy$^{-1}$.

\noindent \textsc{\textbf{System temperature}} --- The receiver system temperature in Kelvin. Throughout this analysis the Parkes MB20 system temperature is taken to be 28 K, as given in the Parkes Telescope Users Guide\footnote{www.parkes.atnf.csiro.au/observing/documentation/user\_guide/}. We note that this is up to 5 K higher than the system temperature used in many publications and that true system temperature also depends on observing elevation.

\noindent \textsc{\textbf{Reference}} --- The journal reference for the burst discovery paper where the event was first reported.

\subsection{Observed parameters}

\noindent \textsc{\textbf{DM}} --- The dispersion measure of the FRB in units of cm$^{-3}$ pc. The integrated electron column density along the line of sight to the burst obtained either with a pulse fitting algorithm or by the search code. For bursts with a published DM produced with a detailed fitting code such as the one described in \citet{Thornton13} this DM is used throughout the re-analysis.

\noindent \textsc{\textbf{DM index}} --- The dispersion measure index of the burst $\alpha$ such that DM $\propto$ $\nu^{-\alpha}$ obtained with a pulse fitting algorithm. The DM index for the propagation of waves through a cold plasma is $\alpha$ = 2 \citep{PulsarHandbook}. 

\noindent \textsc{\textbf{Scattering index}} --- The evolution of pulse width as a function of frequency due to scattering such that W $\propto$ $\nu^{-\beta}$ obtained with a pulse fitting algorithm. The index for the propagation of radio waves through an inhomogeneous turbulent medium is $\beta$ = 4 \citep{PulsarHandbook}. 

\noindent \textsc{\textbf{Scattering time}} --- A measure of the fluctuations in electron density along the line of sight contributing to scattering of the pulse obtained with a pulse fitting algorithm. The number presented in the catalogue is the scattering time for a radio pulse at 1 GHz in ms. Scattering time scales with frequency as $\tau(\nu) = \tau_s (\nu/\nu_0)^{-\beta}$, where $\tau_s$ and $\nu_0$ are at the reference frequency and $\beta$ is the scattering index.

\noindent \textsc{\textbf{Linear polarisation fraction}} --- If polarised data were recoreded for the FRB the fractional linear polarisation is reported with errors. The total linear polarisation is the quadrature sum of Stokes $Q$ and $U$ such that $\sqrt{Q^2+U^2}/I$.

\noindent \textsc{\textbf{Circular polarisation fraction}} --- If polarised data were recorded for the FRB the fractional circular polarisation is reported with errors. The total absolute value of circular polarisation is given by $|V|/I$.

\noindent \textsc{\textbf{S/N}} --- The signal-to-noise of the burst.

\noindent \textsc{\textbf{W$_\textsc{obs}$}} --- The observed width of the FRB in ms obtained either with a pulse fitting algorithm or by the search code. The width reported here is not the \textit{intrinsic} width.

\noindent \textsc{\textbf{\textit{S}$_\textsc{peak,obs}$}} --- The observed peak flux density of the burst in Jy calculated using quantities above via the single pulse radiometer equation \citep{Cordes2003}. Note that this flux density is derived from observed values and is not necessarily the true peak flux density that would be measured if the burst occurred at beam centre; this value should be taken as a lower limit on the true flux density.

\noindent \textsc{\textbf{\textit{F}$_\textsc{obs}$}} --- The observed fluence of the FRB in units of Jy ms calculated as $F_\mathrm{obs} = S_\mathrm{peak,obs} \times W_\mathrm{obs}$. Again, the observed fluence should be taken as a lower limit on the true fluence due to the likely off-axis detection of the burst.

\subsection{Derived parameters}

\noindent \textsc{\textbf{DM$_\textsc{Galaxy}$}} --- The modeled contribution to the FRB DM by the electrons in the Galaxy. The Galactic DM contribution is derived using the NE2001 Galactic electron density model \cite{Cordes02} and should be taken as an estimate as the free electron content of the Galactic halo is not well constrained \cite{Dolag2015}. 

\noindent \textsc{\textbf{DM$_\textsc{excess}$}} --- The DM excess of the FRB over the estimated Galactic DM. This is calculated as $\mathrm{DM}_\mathrm{excess} = \mathrm{DM}_\mathrm{FRB} - \mathrm{DM}_\mathrm{Galaxy}$. 

\noindent \textsc{\textbf{z}} --- The estimated redshift of the FRB based on DM$_\mathrm{excess}$. The redshift is calculated as $z = \mathrm{DM}_\mathrm{excess}/1200$ pc cm$^{-3}$ from estimates of the intergalactic medium (IGM) electron density from \citet{Ioka03}. This relation approximates the full expression to better than 2\% for z < 2. This uncertainty is much less than the line-of-sight variation expected \citep{McQuinn2014}. This value for redshift assumes that any host galaxy or surrounding material contributes nothing to the DM and should be taken as an approximate upper limit on the true redshift.

\noindent \textsc{\textbf{D$_\textsc{comov}$}} --- Comoving distance in units of Gpc derived using the cosmology calculator CosmoCalc \citep{CosmologyCalc}. One should note that this parameter is highly uncertain as, unless an independent redshift measurement is made for an FRB, the comoving distance depends on models of electron density in the Galaxy and IGM as well as the chosen cosmological parameters.

\noindent \textsc{\textbf{D$_\textsc{luminosity}$}} --- Luminosity distance in units of Gpc calculated as $D_L = D_\mathrm{comov} \times (1+z)$. Again, this value is an upper limit based on the upper limit on redshift and comoving distance.

\noindent \textsc{\textbf{Energy}} --- The estimated FRB energy in units of 10$^{32}$ joules calculated as
\begin{equation}
E_\mathrm{FRB} = F_\mathrm{obs} \; \mathrm{BW} \; D_L^2 \times 10^{-29} \; (1+z) \; \mathrm{J}
\end{equation}
\noindent where fluence is in units of Jy ms, bandwidth is in units of Hz, $D_L$ is in units of metres, and $10^{-29}$ is a conversion factor between Jy ms and joules. The equivalent conversion factor from Jy ms to ergs is 10$^{-22}$.

\begin{table*}
\begin{threeparttable}
\caption{Table of select observation parameters (left) and observed parameters (right) from the catalogue. Discovery observations for each burst are listed first. Where a re-analysis of the data has been performed for this work, it is presented as a second entry for the burst. The references are [1] \citet{SarahFRB}, [2] This work, [3] \citet{Keane11}, [4] \citet{Lorimer07}, [5] \citet{Champion2015}, [6] \citet{Thornton13}, [7] (GBT Burst), [8] \citet{Spitler14}, [9] \citet{Ravi2015}, [10] \citet{PetroffFRB}. The FWHM for the telescopes include in this table are 15$'$ (Parkes), 7$'$ (Arecibo), and 16$'$ (GBT). Question marks denote values that were not speicified in the original publication or were not available publicly. }
\begin{center}
\setlength{\extrarowheight}{3 pt}
\begin{tabular}{lccc|cccccc}
\hline\hline
FRB name & Telescope & $gl^\textbf{a}$ & $gb^\textbf{a}$ & DM & S/N & W$_\mathrm{obs}$ & $S_\mathrm{peak,obs}$ & $F_\mathrm{obs}$ & Ref. \\
	& & (deg) & (deg) & (pc cm$^{-3}$) & & (ms) & (Jy) & (Jy ms) & \\
\hline%
FRB 010125$^{\textbf{b}}$ & Parkes & 356.641 & -20.021 & 790(3) & 17 & 9.4(2) & 0.3 & 2.82 & [1] \\
 & & & & 790(2) & 25 & 10.6$^{+2.8}_{-2.5}$ & 0.54$^{+0.11}_{-0.07}$ & 5.7$^{+2.9}_{1.9}$ & [2] \\
FRB 010621 & Parkes & 25.434(3) & -4.004(3) & 745(10) & -- & 7 & 0.410 & 2.870 & [3] \\
  &  & & & 748(3) & 18 & 8$^{+4.0}_{-2.3}$ & 0.53$^{+0.26}_{-0.09}$ & 4.2$^{+5.2}_{-1.7}$ & [2] \\
FRB 010724 & Parkes & 300.653(3) & -41.805(3) & 375 & >23 & 5.0 & 30(10) & 150 & [4] \\
  & & & & 375(2) & >100 & >20 & >1.57 & >31.4 & [2] \\
  & & 300.913(3) & -42.427(3) & -- & -- & -- & -- & -- & [4] \\
  & & & & 375(2) & 16 & 13.0$^{+5.0}_{-11.0}$ & 0.29$^{+0.18}_{-0.09}$ & 3.7$^{+4.7}_{-3.4}$ & [2] \\
  & & 301.310(3) & -41.831(3) & -- & -- & -- & -- & -- & [4] \\
  & & & & 375(2) & 26 & 16$^{+5}_{-4}$ & 0.54$^{+0.14}_{-0.07}$ & 8.6$^{+5.6}_{-3.0}$ & [2] \\
  & & 300.367(3) & -41.368(3) & 375(2) & 6 & 33$^{+12}_{-28}$ & 0.09$^{+0.03}_{-0.04}$ & 2.9$^{+2.4}_{-2.6}$ & [2] \\
FRB 090625 & Parkes & 226.444(3) & -60.030(3) & 899.6(1) & 28 & -- & -- & 2.2 & [5] \\
  & & & & 899.6(1) & 30 & 1.9$^{+0.8}_{-0.7}$ & 1.14$^{0.42}_{-0.21}$ & 2.2$^{+2.1}_{-1.1}$ & [2] \\
FRB 110220 & Parkes & 50.829(3) & -54.766(3) & 944.38(5) & 49 & 5.6(1) & 1.3 & 7.3(1) & [6] \\
  &  & & & 944.38(5) & 54 & 6.6$^{+1.3}_{-1.0}$ & 1.1$^{+0.2}_{-0.1}$ & 7.3$^{+2.6}_{-1.7}$ & [2] \\
FRB 110523 & GBT & 56.12(?) & -37.82(?) & 623.30(6) & 42 & 1.73(17) & 0.6 & -- & [7] \\
FRB 110626$^{\textbf{c}}$ & Parkes & 355.862(3) & -41.752(3) & 723.0(3) & 11 & 1.4 & 0.4 & 0.56 & [6] \\
  &  & & & 723.0(3) & 12 & 1.4$^{+1.2}_{-0.5}$ & 0.6$^{+1.2}_{-0.1}$ & 0.9$^{+4.0}_{-0.4}$ & [2] \\
FRB 110703 & Parkes & 80.998(3) & -59.019(3) & 1103.6(7) & 16 & 4.3 & 0.5 & 2.15 & [6] \\
  &  & & & 1103.6(7) & 17 & 3.9$^{+2.2}_{-1.9}$ & 0.45$^{+0.28}_{-0.10}$ & 1.7$^{+2.7}_{-1.0}$ & [2] \\
FRB 120127 & Parkes & 49.287(3) & -66.204(3) & 553.3(3) & 11 & 1.1 & 0.5 & 0.55 & [6] \\
  &  & & & 553.3(3) & 13 & 1.2$^{+0.6}_{-0.3}$ & 0.6$^{+0.4}_{-0.1}$ & 0.7$^{+1.0}_{-0.3}$ & [2] \\
FRB 121002 & Parkes & 308.220(3) & -26.265(3) & 1629.18(2) & 16 & -- & -- & 2.3 & [5] \\
 & & & & 1629.18(2) & 16 & 5.4$^{+3.5}_{-1.2}$ & 0.43$^{+0.33}_{-0.06}$ & 2.3$^{+4.5}_{-0.8}$ & [2] \\
FRB 121102 & Arecibo & 174.950(2) & -0.225(2) & 557(2) & 14 & 3.0(5) & 0.4$^{+0.4}_{-0.1}$ & 1.2$^{+1.6}_{-0.5}$ & [8] \\
FRB 130626 & Parkes & 7.450(3) & 27.420(3) & 952.4(1) & 20 & -- & -- & 1.5 & [5] \\
 & & & & 952.4(1) & 21 & 1.98$^{+1.2}_{-0.44}$ & 0.74$^{+0.49}_{-0.11}$ & 1.5$^{+2.5}_{-0.5}$ & [2] \\
FRB 130628 & Parkes & 225.955(3) & 30.656(3) & 469.88(1) & 29 & & & 1.2 & [5]\\
 & & & & 469.88(1) & 29 & 0.64(13) & 1.9$^{+0.3}_{-0.2}$ & 1.2$^{+0.5}_{-0.4}$ & [2] \\
FRB 130729 & Parkes & 324.788(3) & 54.745(3) & 861(2) & 14 & -- & -- & 3.5 & [5] \\
 & & & & 861(2) & 14 & 15.6$^{+9.9}_{-6.2}$ & 0.22$^{+0.17}_{-0.05}$ & 3.4$^{+6.5}_{-1.8}$ & [2] \\
FRB 131104 & Parkes & 260.466(3) & -21.839(3) & 779(1) & 30 & 2.08 & 1.12 & 2.33 & [9] \\
  &  & & & 779(2) & 34 & 2.4$^{+0.9}_{-0.5}$ & 1.16$^{+0.35}_{-0.13}$ & 2.8$^{+2.2}_{-0.8}$ & [2] \\
FRB 140514 & Parkes & 50.841(3) & -54.612(3) & 562.7(6) & 16 & 2.8$^{+3.5}_{0.7}$ & 0.47$^{+0.11}_{-0.08}$ & 1.3$^{+2.3}_{-0.5}$ & [10] \\
  &  & & & 562.7(6) & 16 & 2.82$^{+0.64}_{-2.11}$ & 0.47$^{+0.10}_{-0.14}$ & 1.3$^{+0.6}_{-1.1}$ & [2] \\
\hline\hline
\end{tabular}
\begin{tablenotes}
\small 
\item $^\textbf{a}$ Errors in $gl$ and $gb$ refer to the pointing accuracy of the given telescope.
\item $^\textbf{b}$ Originally incorrectly published as FRB 011025.
\item $^\textbf{c}$ Originally incorrectly published as FRB 110627.
\end{tablenotes}
\end{center}
\label{tab:tab1}
\end{threeparttable}
\end{table*}

\section{HOW TO USE THE CATALOGUE}
\label{sec:howto}

The catalogue can be viewed on the web or downloaded as a plain text file for use offline. The homepage of the catalogue\footnote{http://www.astronomy.swin.edu.au/pulsar/frbcat/} presents all the FRBs with a few key parameters: telescope name, $gl$, $gb$, FWHM, DM, S/N, W$_\mathrm{obs}$, $S_\mathrm{peak,obs}$, $F_\mathrm{obs}$, and reference. For each burst there is a detailed page containing all the parameters described in Section~\ref{sec:quantities}. Individual pages for the bursts contain relevant images and figures such as the dispersion sweep for the burst, polarisation (if available). Images are also available for download from the catalogue page. 

\begin{figure*}
\centering
\includegraphics[height=22cm]{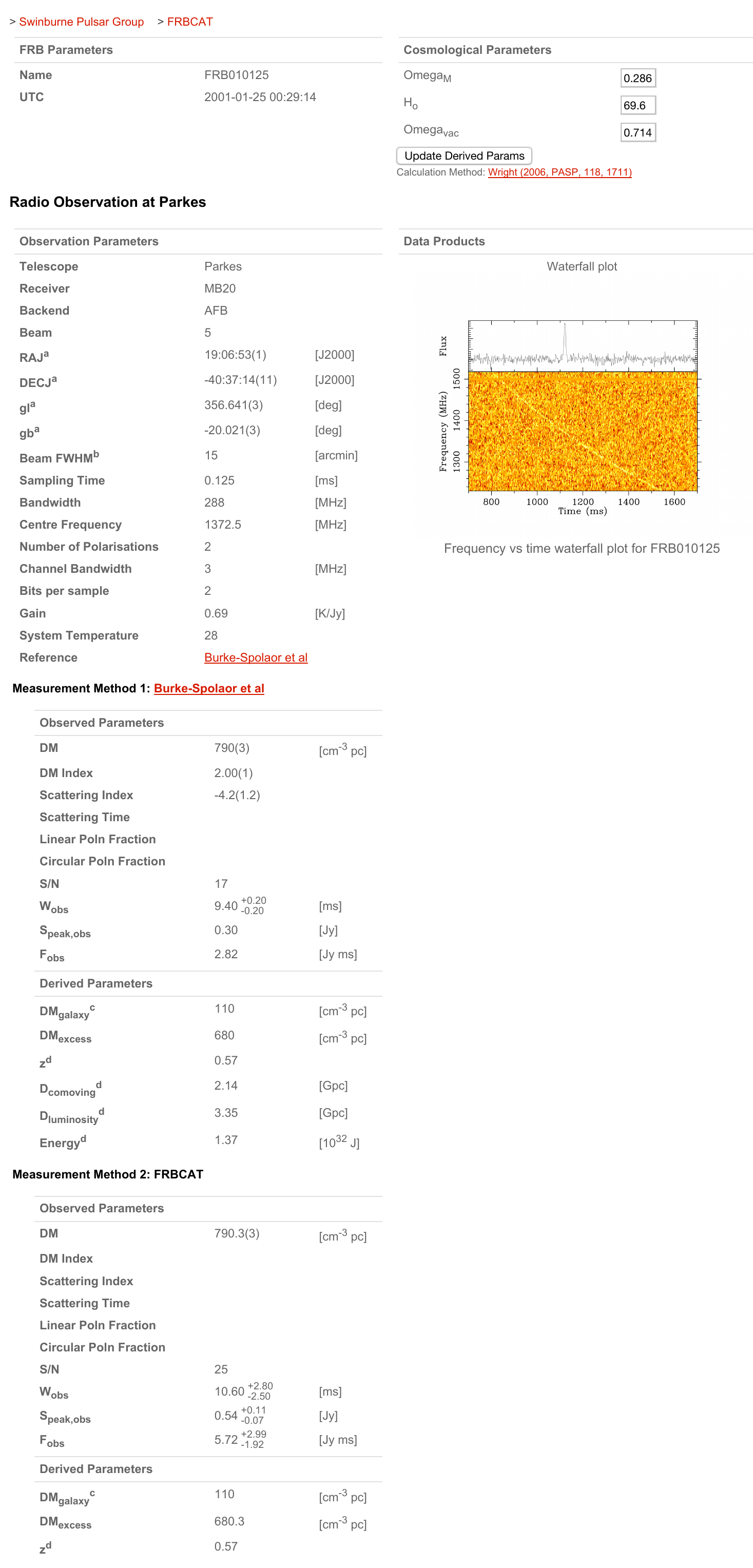}
\caption{Example of the beginning of an FRB entry on the catalogue webpage. Telescope-specific observation parameters have been separated from the observed parameters measured from the available data. Where the data have been re-analysed for the catalogue multiple measurement methods are available. \label{fig:cosmocalc}}
\end{figure*}

Cosmological parameters such as $D_\mathrm{comov}$, $D_\mathrm{luminosity}$, and Energy are derived on-the-fly using the Cosmology Calculator module from \citet{CosmologyCalc} with the input parameters displayed on the individual burst sub-pages. Inputs for $\Omega_\mathrm{M}$, $\Omega_\Lambda$, and $H_0$ can be modified on the catalogue webpage; default values are $\Omega_\mathrm{M}$ = 0.286, $\Omega_\mathrm{vac}$ = 0.714, and $H_0$ = 69.6, Figure~\ref{fig:cosmocalc}. Updating these numbers will automatically update the calculated values of comoving distance, luminosity distance, and energy. 

Alternatively, the catalogue can be downloaded either in CSV or tabular plain-text format from the catalogue homepage. The file generated will have all information contained in the most recent version of the online catalogue. If derived and measured parameters for a burst have been derived using multiple methods (i.e. values from publication and re-analysis) all methods will be included in the downloaded table with the appropriate reference. The value used in any studies of the bursts is the choice of the user.

\section{CLOSING REMARKS}
\label{sec:conclusions}

In this paper we present the FRBCAT, an online catalogue of fast radio bursts. This catalogue includes all bursts currently available in the literature and will be updated as new bursts are published in the future. The catalogue presents an overview of all bursts on the main page but also includes a page for each individual burst with a number of parameters that describe the observational setup, the observed burst properties, and model-dependent cosmological parameters. Additionally, all bursts for which the data are public have been re-analysed using a standardised method in an effort to make detections more consistent and directly comparable. The tools for our re-analysis have been made available through github and the data can be processed with freely available software packages and processing tools. It is our hope that data for all future bursts will be made public upon publication and this catalogue will encourage communication and collaboration.

\begin{acknowledgements}
The Parkes radio telescope and the Australia Telescope Compact Array are part of the Australia Telescope National Facility which is funded by the Commonwealth of Australia for operation as a National Facility managed by CSIRO. Parts of this research were conducted by the Australian Research Council Centre of Excellence for All-sky Astrophysics (CAASTRO), through project number CE110001020. This work was performed on the gSTAR national facility at Swinburne University of Technology. gSTAR is funded by Swinburne and the Australian Government's Education Investment Fund. The authors would like to thank the SUPERB\footnote{https://sites.google.com/site/publicsuperb/} collaboration for their help in beta testing the website.
\end{acknowledgements}

\bibliographystyle{apj}
\bibliography{FRBthesis}

\end{document}